# Liquid-liquid transition revealed by quasi-static cooling of an ultra-viscous metallic liquid

S. Hechler[1,2], B. Ruta[2,3*], M. Stolpe[1], E. Pineda[4], Z. Evenson[5], O. Gross[1], W. Hembree[1], A. Bernasconi[2], R. Busch[1] and I. Gallino[1]

[1] *Chair of Metallic Materials, Department of Materials Science and Engineering, Saarland University, Campus C6.3, 66123 Saarbrücken, Germany*

[2] *ESRF—The European Synchrotron, CS40220, 38043 Grenoble, France*

[3] *Institut Lumière Matière, UMR5306 Université Lyon 1-CNRS, Université de Lyon, 69622 Villeurbanne Cedex, France*

[4] *Departament de Física, Universitat Politècnica de Catalunya - BarcelonaTech, ESAB, Esteve Terradas 8, 08860 Castelldefels, Spain*

[5] *Heinz Maier-Leibnitz-Zentrum (MLZ) and Physik Department, Technische Universität München, Lichtenbergstrasse 1, 85748 Garching, Germany*

*Corresponding author: beatrice.ruta@univ-lyon1.fr

**Abstract**

Temperature-driven polyamorphism has been reported in various supercooled liquids and glasses. The dynamical and structural routes followed by the system during such crossovers are however not universal and appear to be related to intrinsic kinetic properties of the liquid. By combining x-ray photon correlation spectroscopy and high energy x-ray diffraction, we have followed the collective atomic motion and structural changes during quasi-static cooling of the $Au_{49}Cu_{26.9}Si_{16.3}Ag_{5.5}Pd_{2.3}$ metallic glass-former. Due to this ultra-slow thermal protocol, the glass transition temperature is lowered far enough to reveal a liquid-liquid crossover in the ultra-viscous supercooled phase. This transition is usually hidden by vitrification and leads to a strong liquid characterized by increasing correlation at the medium range.

**Article:**

Polyamorphic liquid-liquid transitions (LLT) have been found in a large variety of glass-forming liquids in both experiments[1–8] and simulations[9–11]. When observed upon cooling, dynamical quantities like viscosity and structural relaxation time can show a fragile-to-strong crossover. Within the kinetic fragility concept[12–14], the liquid evolves, during the LLT, from a high temperature fragile phase with a steep temperature dependence of the dynamics, to a strong phase less affected by temperature changes. For fragile liquids, such as molecular and some metallic glass-formers, the LLT is expected at such a low temperature that the system may arrest beforehand[15]. However, no direct experimental evidence of this has been reported so far.

Here, we observe a fragile-to-strong LLT in a $Au_{49}Cu_{26.9}Si_{16.3}Ag_{5.5}Pd_{2.3}$ metallic glass-former by investigating the microscopic structural changes and collective atomic-scale dynamics upon cooling in the ultra-viscous regime. By applying a quasi-static cooling from the supercooled liquid, $T_g$ is lowered more than 30 degrees in comparison to a standard cooling rate of 20 K

min$^{-1}$ (from ref.$^{16}$, see also SI), allowing us to observe the fragile-to-strong crossover in an ultra-slow metallic glass-forming liquid with structural relaxation times of some thousands of seconds.

Dynamical and structural measurements were performed by applying an identical thermal protocol using x-ray photon correlation spectroscopy (XPCS) and high energy x-ray diffraction (XRD) at beamlines ID10 and ID11 at ESRF, France. Glassy ribbons were heated from the as-spun state into the supercooled liquid with an isothermal stepwise heating protocol up to 396 K, and subsequently cooled down to 380 K with a cooling protocol consisting of isothermal steps of 0.5 K, at a cooling rate of 0.1 K min$^{-1}$. Each isothermal step last 5-40 times longer than the measured relaxation time, assuring thus the equilibrium condition. Within this extremely slow protocol, the system equilibrated to the new temperature before the beginning of the measurements at all temperatures but 380 K. Here, the equilibration was monitored *in-situ*. The XRD data are further compared to additional measurements performed at beamline P02.1 at DESY, Germany, with a faster cooling rate of 1.5 K min$^{-1}$. Structural relaxation times measured with a dynamic mechanical analyzer (DMA) at high temperature in the supercooled liquid are also reported for comparison. Details of the different experimental methodologies can be found in the SI.

Two main quantities are accessible with XPCS, the two-time correlation function (TTCF) which describes the instantaneous correlation between intensity fluctuations in subsequent speckles patterns, and its temporal average given by the intensity autocorrelation function $g_2(Q,t)$[17,18]. The TTCF shows the temporal evolution of the microscopic dynamics during the measurement, while $g_2(Q,t)$ is connected to the intermediate scattering function F(Q,t) and provides quantitative information on the structural relaxation time at the probed length scale and -time interval[19].

Figure 1 shows XPCS data measured at a wave vector $Q_p$=2.78 Å$^{-1}$ corresponding to the maximum of the static structure factor, S(Q), for different isothermal steps between 392 K and 380 K while cooling from the supercooled liquid. These temperatures lie slightly above the T$_g$ of the applied cooling rate, which is determined to be 380 K, thus 33 K below the standard calorimetric $T_g^{end}$ = 413 K (from ref. [16], see SI). The time average intensity auto-correlation functions are reported in Fig. 1a, together with the best fits obtained using a Kohlrausch-Williams-Watts (KWW) model function: $g_2(Q,t) = 1 + c[\exp(-2(t/\tau)^\beta]$, where $\tau$ is the structural relaxation time and $\beta$ the shape parameter. The data are normalized by the parameter *c*, which is the product of the experimental contrast and nonergodicity factor of the glass and is found to be ~3% at all investigated temperatures. Despite the narrow range of investigated temperatures, the decay of the curves shifts dramatically toward longer times as a function of decreasing temperature. This corresponds to a dramatic increase in the intrinsic relaxation time $\tau$ of two orders of magnitude (~10$^2$ s to ~10$^4$ s) in a span of just 12 degrees. This behavior reflects the steep temperature dependence of the supercooled liquid's relaxation kinetics in the vicinity of T$_g$ and it is associated to a heterogeneous relaxation spectrum[20–24]. As expected for supercooled liquids, all curves indeed decay in a stretched exponential fashion with <β> ≈ 0.87 ± 0.10 and follow the typical time-temperature superposition (see inset of Fig. 1a)[25–28].

The stretched exponential shape of the intensity autocorrelation functions in the supercooled liquid contrasts dramatically with its behavior in the glass. Metallic glasses are indeed known to exhibit a stress-dominated atomic dynamics characterized by an anomalous compressed decay of the correlation functions (i.e. with β > 1)[23,25,29,30], reminiscent of that reported in some soft glasses[20,31], likely related to microscopic elastic frustrations[32–34]. As an example, Fig. 1b shows two normalized $g_2(Q_p,t)$ functions as a function of $t/\tau$, measured at the same temperature of 383 K, one in the supercooled liquid state (open orange diamonds) and the other in the glassy state (blue open circles) The dashed-dotted line corresponds to a simple exponential decay for comparison. A similar dynamical crossover has also been reported in a Mg-based metallic glass while cooling from the liquid into the glass and vice versa [30,35]. It is important to note here that the two curves in Fig. 1b correspond to dynamics that differ by about a factor of 15 in timescales, with $\tau$ values of only ~200 s in the glass and of ~3000 s in the ultra-viscous liquid. The equilibrium nature of the supercooled liquid is illustrated also in Fig. 1c, by the constant width of the intensity profile along the main diagonal in the TTCF measured at 386 K. Here, the atomic dynamics appears stationary with no sign of aging over the entire measuring time of 9000 s, thus for a temporal interval ~8 times larger than the corresponding structural relaxation time. This applies to all $T$ but $T_g$ (i.e. 380 K for the quasi-static cooling) whose corresponding TTCF is reported in Fig. 2. Within this protocol, the system is not able to follow the temperature changes and temporarily freezes into the glassy state from which it equilibrates in about 2800 s, as signaled by the continuous broadening of the intensity profile at the beginning of the measurement. After equilibration, the dynamics becomes stationary again and the system reach the liquid phase. The corresponding $g_2(Q_p,t)$ at 380 K after equilibration is shown in Fig. 1a. It should be noted that the observed equilibration is highly promoted by temperature down-jumps. This is indeed also the reason why the system can follow the supercooled behavior during the quasi-static cooling. On the other hand, as shown in previous XPCS studies, the spontaneous temporal equilibration towards the liquid phase is strongly impeded by temperature up-jumps from the glassy state[23]. This asymmetric equilibration is typical of glass forming systems[36] and the magnitude of such an effect appears to be dramatically large at the atomic level.

Figure 3a shows the temperature dependence of the mean relaxation time $\langle\tau\rangle = \Gamma\left(\frac{1}{\beta}\right)\frac{\tau}{\beta}$ obtained from the fits of the XPCS data within the entire quasi-static cooling protocol. A clear change in the trend is observed at 389 K where $\langle\tau\rangle$ evolves from a steep to a shallower temperature dependence as also highlighted in the inset for better clarity. This temperature is about 10 K above $T_g$ of the quasi-static cooling rate, and it is about 20 K below the calorimetric $T_g^{end}$ (defined at 20 K min$^{-1}$ taken from [16]). The data above and below 389 K can be fitted separately with a Vogel-Tamman-Fulcher (VFT) equation of the type $\langle\tau\rangle = \tau_0 \cdot exp\left(\frac{D^* \cdot T_0}{T - T_0}\right)$, where $\tau_0$ is the high temperature limit of the relaxation time (10$^{-14}$ s), $D^*$ is the kinetic fragility parameter and $T_0$ is the VFT-temperature. The high-temperature liquid exhibits a fragile behavior with $D^* = 8.9 \pm 0.4$ and $T_0 = 315.9 \pm 2.3$ K. This fit is in good agreement with the fragility obtained at the higher temperatures of the DMA α-relaxation times (triangles in Fig. 3a). However, the low-temperature liquid is best fitted with $D^* = 23.1 \pm 0.8$ and $T_0 = 243.3 \pm 3$ K, suggesting a clearly stronger kinetic nature of the liquid and, hence, a well-defined liquid-liquid transition in the ultra-viscous state. The value of $D^*$ obtained here for the strong liquid is indeed comparable to that of the strongest metallic glass formers[37] in the vicinity of $T_g$, whereas that obtained for the fragile liquid is even somewhat smaller than what has been reported for the most fragile bulk metallic glass formers based on Pd or Pt[32,33]. In Zr-based metallic glass-forming systems, for example, such a fragile behavior is typically detected in the stable liquid well above the LLT critical temperature[3,5,38].

The LLT reported here is also accompanied by peculiar structural changes as shown in Fig. 3b where we report the temperature dependence of the position of the first sharp diffraction peak (FSDP) of the total static structure factor as $(Q_p(T_{ref})/Q_p(T))^3$, with $T_{ref}$=385.5 K. In the two datasets presented, the green triangles are taken by continuously cooling the liquid with 1.5 K min$^{-1}$ and exhibit the typical behavior associated to the glass transition: at the corresponding $T_g$ of 390 K, $(Q_p(T_{ref})/Q_p(T))^3$ bends over to a weaker temperature dependence due to the kinetic arrest and transition into the glassy state. By applying the same quasi-static cooling used in the XPCS experiments, $T_g$ is lowered to 380 K (see SI). Normally, one would simply expect to observe the departure from the liquid behavior at this lower temperature (see SI). In stark contrast to this, however, $(Q_p(T_{ref})/Q_p(T))^3$ (blue diamonds) displays an anomalous increase between 395.5 and 385.5 K, i.e. above the expected $T_g$, and then decreases again at lower temperatures. Below 385.5 K the data are obtained upon cooling with a rate of 7 K min$^{-1}$ interrupted by an isothermal step at 380 K to investigate the structural equilibration in-situ.

The steady increase upon cooling in $(Q_p(T_{ref})/Q_p(T))^3$ can be viewed as a signature of growing spatial correlations at the medium range[39]. It cannot be associated to vitrification and thus strengthens the case of a liquid-liquid transition. Indeed, not only does it occur in proximity to the dynamical transition observed in XPCS, but it also strongly resembles what has been reported during a LLT in a Zr-based metallic liquid far above $T_g$[5].

Interestingly, upon cooling, the structure of the fragile high-temperature liquid evolves steadily towards the strong low temperature liquid within a temperature interval of 10 K, whereas the dynamical crossover appears to occur at a well-defined temperature. This is in good agreement with other results where sudden dynamical changes are accompanied by smeared-out structural changes[30]. After the in-situ equilibration at 380 K, the strong liquid is vitrified by continuous cooling. It isimportant to note that the LLT is not observed when the system is cooled with a faster rate. As shown by the green triangles in Fig. 3b, in this case, $T_g$ is high enough that the fragile liquid freezes into the glassy state before the LLT can occur.

Although we observe the dynamics as stationary in the equilibrium supercooled liquid, this is, however, not the case for the structure. The corresponding temporal evolution during the isothermal holds is shown in Fig. 4 for both $(Q_p(T_{ref})/Q_p(T))^3$ (panel a) and $\Gamma/\Gamma_0(T_{ref})$, the relative change of the full width at half maximum (FWHM) of the FSDP (panel b). Both quantities exhibit minute increases at temperatures below 390 K, whereas there are no visible changes above this temperature. This difference is highlighted in the inset for 389 K and 391 K. Only at 380 K do the data decrease with time. This latter behavior is associated to structural relaxation of the glass during equilibration. As shown in Fig. 2, at this low temperature, the system falls in the glassy state and equilibrates within ~2800 s. The same applies for the structural metrics in Fig. 3 (see also SI) and can be attributed to a densification and medium range ordering during aging[40]. The behavior at the other temperatures appears anomalous, as one would naturally expect a stable structure in the supercooled liquid, in agreement also with the stationary dynamics. These data, on the other hand, show a small, but significant, structural evolution in the crossover region below 390 K. To the best of our knowledge, this is the first time that such a kind of a temporal structural evolution is reported, as usually polyamorphism is investigated by looking at temperature and pressure dependencies rather than temporal ones.

This behavior could be a consequence of the sluggish dynamics that impedes fast structural rearrangements. The fact that the relaxation time keeps constant means that the ongoing structural rearrangements do not imply average density changes, as indeed is often the case during high temperature LLTs[5,38]. A slow crystallization could also lead to structural changes without affecting the dynamics. In this case, crystal growth would appear as an increase of the static contribution in the XPCS spectra. However, we can rule out this possibility as we do not find any evidence for crystallization neither in the XPCS and XRD data, nor in calorimetric measurements performed after the entire thermal treatment (see SI). Finally, it is worth mentioning that a similar phenomenology has been observed also in the stationary regime of aging found at low temperatures in metallic glasses[40]. Here the aging occurs with jumps through stationary regimes of dynamics[29], where however the structure keeps changing, signaling increasing correlations at the medium range[40].

In conclusion, our study reports on the collective atomic motion and structural changes occurring at the atomic level in a $Au_{49}Cu_{26.9}Si_{16.3}Ag_{5.5}Pd_{2.3}$ metallic glass-forming liquid. By applying a quasi-static cooling, $T_g$ is lowered by more than 30 K and the system undergoes a fragile-to-strong liquid-liquid transition which is usually hidden by vitrification at faster cooling rates. This dynamical crossover is accompanied by anomalous structural changes at the same length scale, suggesting reorganizations at the medium range. Differently from the dynamics, the S(Q) evolves spontaneously with time, reflecting a complex phenomenology. A fundamental question is whether similar LLTs could occur also in other fragile glass formers as it was suggested[15] or even in all glass forming liquids. The complex phenomenology here discussed clearly calls for further experimental and theoretical investigations

We gratefully thank the ESRF and DESY for providing beamtime. H. Vitoux, K. L'Hoste and F. Zontone are acknowledged for the support during the XPCS measurements at ESRF, Y. Chushkin for providing the code for the analysis of the XPCS data, and J. Bednarcik for the support during the XRD measurements at DESY. I. Gallino acknowledges financial support from the DFG Grant No. GA 1721/2-2. E. Pineda acknowledges financial support from MINECO, grant FIS2014-54734-P, and Generalitat de Catalunya, grant 2014SGR00581.

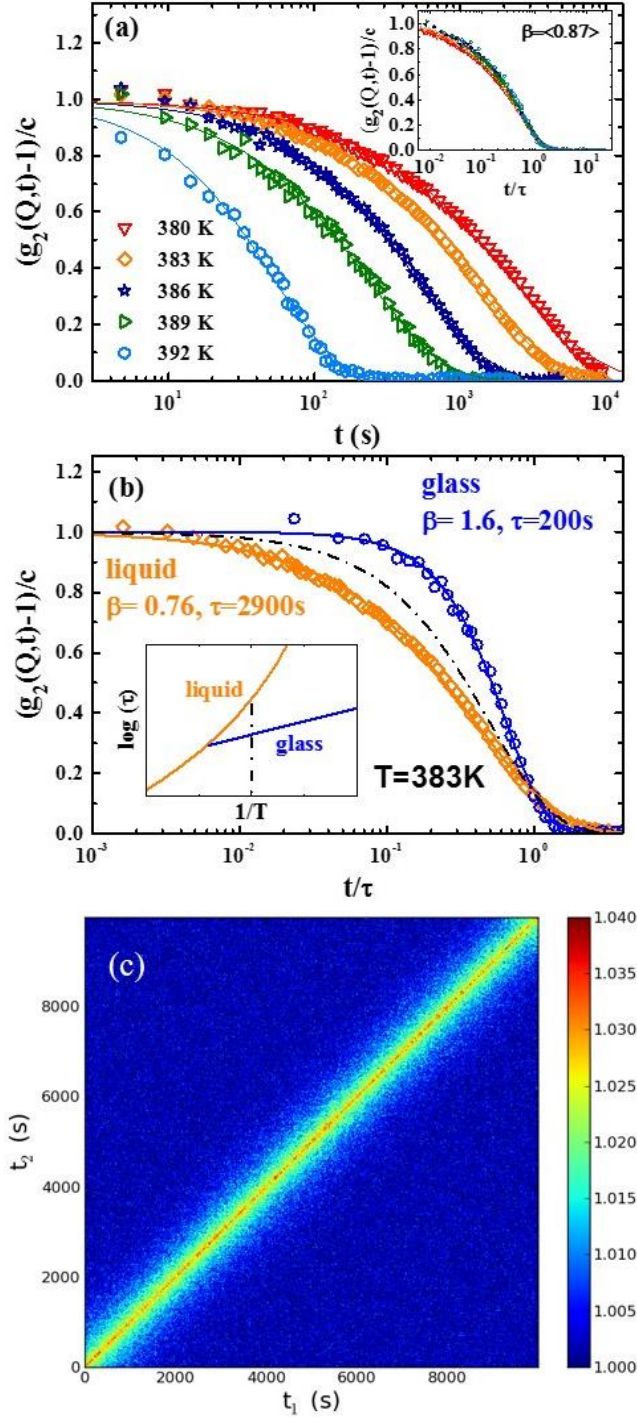

**Figure 1:** **(a)** Temperature dependence of normalized intensity autocorrelation functions measured with XPCS at $Q_p$=2.78 Å$^{-1}$. Lines are fits using the KWW function. **Inset**: Same data reported as $t/\tau$. **(b)** $g_2(Q_p,t)-1/c$ as a function of the rescaled time measured at 383 K in an hyperquenched glass heated from low temperature (blue circles), and in the supercooled liquid shown in panel (a) (orange diamonds). The dashed line is a single exponential decay. The two curves correspond to two distinct dynamics as sketched in the inset by the intersection of the vertical line with the glass or the liquid behaviors. **(c)** TTCF at 386 K showing stationary dynamics over the whole probed interval.

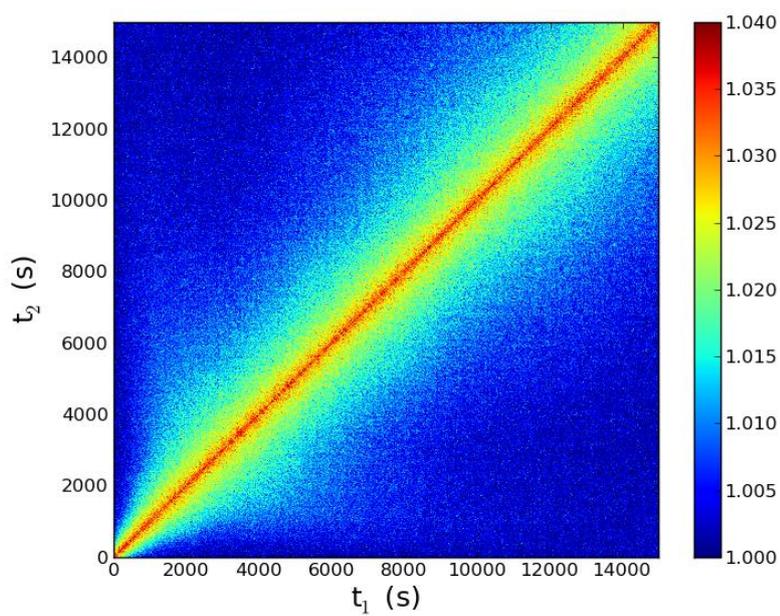

**Figure 2**: TTCF measured with XPCS at 380 K after quenching from 385.5 K with 7 K min$^{-1}$. The broadening at short times due to equilibration stops after ~2800 s.

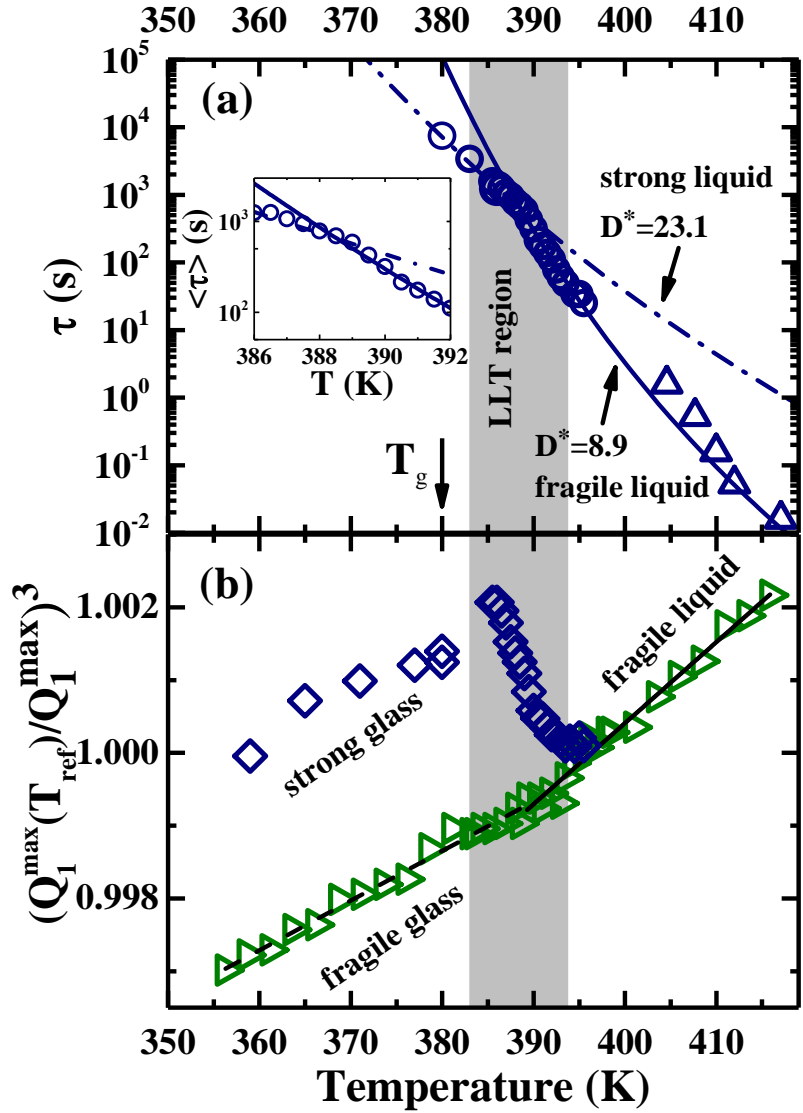

**Figure 3:(a)** Temperature dependence of the relaxation time measured by XPCS (circles) and DMA (triangles). The LLT occurs at 389 K leading to two separates regimes with distinct fragilities as quantified by VFT fits. The transition region is magnified in the inset. **(b)** Temperature dependence of the relative shift $(Q_1^{max}(T_{ref})/Q_1^{max})^3$ of the FSDP measured with XRD by continuous cooling with 1.5 K min$^{-1}$ (green triangles), and by applying the same quasi-static protocol used for the XPCS data (blue diamonds).

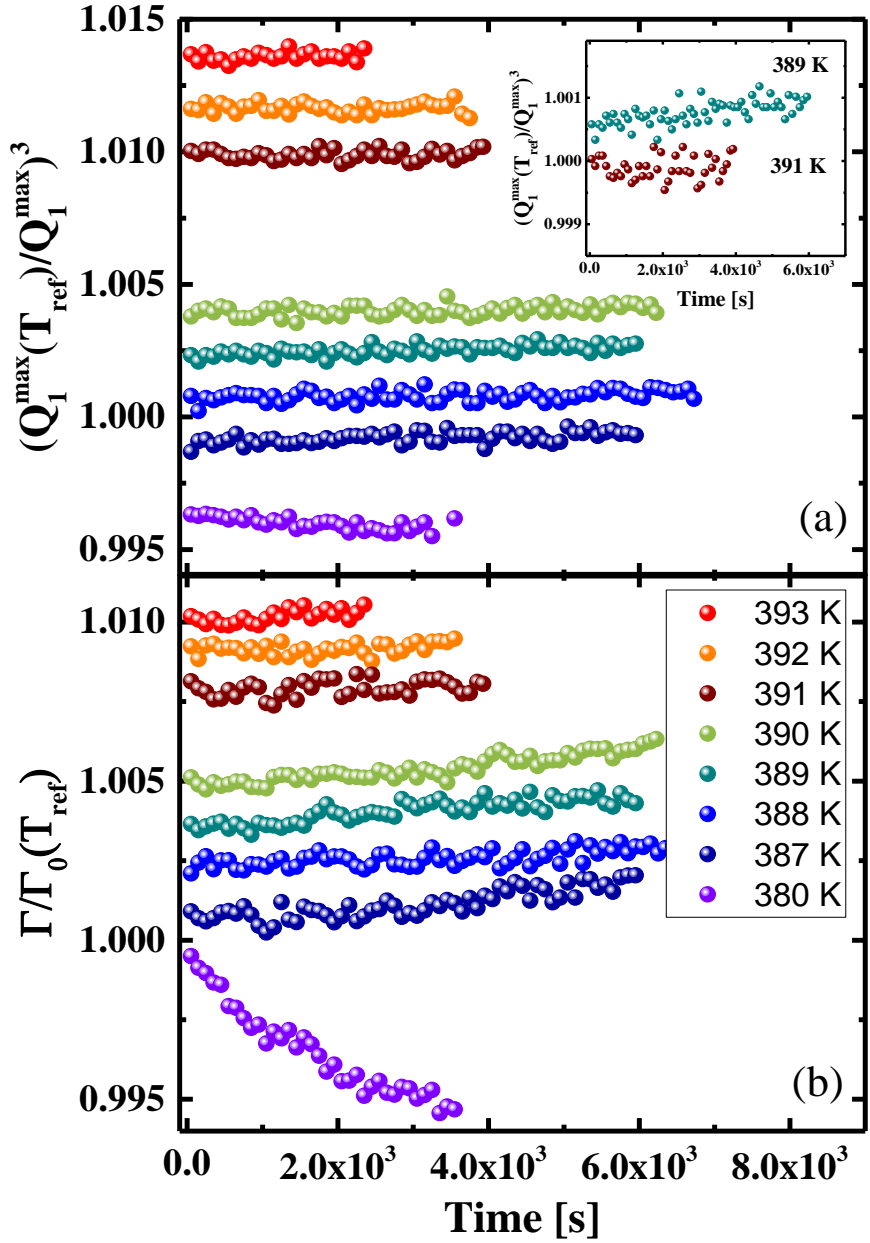

**Figure 4:** Temporal evolution of $(Q_1^{max}(T_{ref})/Q_1^{max})^3$ **(a)** and FWHM **(b)** during isotherms in the transition region between 395.5 and 385.5 K. All data are arbitrary vertically shifted for clarity. Both quantities slightly increase with time at all temperatures but 380 K where they decrease due to the equilibration at short times from the glassy state.

**References**

1. Aasland, S. & McMillan, P. F. Density-driven liquid-liquid phase separation in the system Al2O3-Y2O3. *Lett. to NatureNature* **369,** 633–636 (1994).

2. Katayama, Y. *et al.* Macroscopic Separation of Dense Fluid Phase and Liquid Phase of Phosphorus. *Sci. Reports* **306,** 848–851 (2004).

3. Way, C., Wadhwa, P. & Busch, R. The influence of shear rate and temperature on the viscosity and fragility of the Zr41.2Ti13.8Cu12.5Ni10.0Be22.5 metallic-glass-forming liquid. *Acta Mater.* **55,** 2977–2983 (2007).

4. Evenson, Z., Schmitt, T., Nicola, M., Gallino, I. & Busch, R. High temperature melt viscosity and fragile to strong transition in Zr-Cu-Ni-Al-Nb(Ti) and Cu 47Ti 34Zr 11Ni 8 bulk metallic glasses. *Acta Mater.* **60,** 4712–4719 (2012).

5. Wei, S. *et al.* Liquid-liquid transition in a strong bulk metallic glass-forming liquid. *Nat. Commun.* **4,** 2083 (2013).

6. Kobayashi, M. & Tanaka, H. The reversibility and first-order nature of liquid-liquid transition in a molecular liquid. *Nat. Commun.* **7,** 13438 (2016).

7. Gallo, P. *et al.* Water: A Tale of Two Liquids. *Chem. Rev.* **116,** 7463–7500 (2016).

8. Katayama, Y. *et al.* A first-order liquid-liquid phase transition in phosphorus. *Nature* **403,** 170–173 (2000).

9. Saika-Voivod, I., Poole, P. H. & Sciortino, F. Fragile-to-strong transition and polyamorphism in the energy landscape of liquid silica. *Nature* **412,** 514–517 (2001).

10. Molinero, V., Sastry, S. & Angell, C. A. Tuning of tetrahedrality in a silicon potential yields a series of monatomic (Metal-like) glass formers of very high fragility. *Phys. Rev. Lett.* **97,** 1–4 (2006).

11. Bhat, M. H. *et al.* Vitrification of a monatomic metallic liquid. *Nature* **448,** 787–90 (2007).

12. Angell, C. A. Relaxation in liquids, polymers and plastic crystals — strong/fragile patterns and problems. *J. Non. Cryst. Solids* **131,** 13–31 (1991).

13. Böhmer, R., Ngai, K. L., Angell, C. A. & Plazek, D. J. Nonexponential relaxations in strong and fragile glass formers. *J. Chem. Phys.* **99,** 4201 (1993).

14. Angell, C. A. Formation of Glasses from Liquids and Biopolymers. *Science (80-. ).* **267,** 1924–1935 (1995).

15. Angell, C. A. Glass-Formers and Viscous Liquid Slowdown since David Turnbull: Enduring Puzzles and New Twists. *MRS Bull.* **33,** 544–555 (2011).

16. Evenson, Z. *et al.* β relaxation and low-temperature aging in a Au-based bulk metallic glass: From elastic properties to atomic-scale structure. *Phys. Rev. B* **89,** 174204 (2014).

17. Madsen, A., Leheny, R. L., Guo, H., Sprung, M. & Czakkel, O. Beyond simple exponential correlation functions and equilibrium dynamics in x-ray photon correlation spectroscopy. *New J. Phys.* **12,** 55001 (2010).

18. Chushkin, Y., Caronna, C. & Madsen, A. A novel event correlation scheme for X-ray photon correlation spectroscopy. *J. Appl. Crystallogr.* **45,** 807–813 (2012).

19. Madsen, A., Fluerasu, A. & Ruta, B. in *Synchrotron Light Sources and Free-Electron Lasers* 1–21 (Springer International Publishing, 2015). doi:10.1007/978-3-319-04507-8_29-1

20. Cipelletti, L. *et al.* Universal non-diffusive slow dynamics in aging soft matter. *Faraday Discuss.* **123,** 237-251-322, 419–421 (2003).

21. Cipelletti, L., Manley, S., Ball, R. C. & Weitz, D. a. Universal Aging Features in the Restructuring of Fractal Colloidal Gels. *Phys. Rev. Lett.* **84,** 2275–2278 (2000).

22. Berthier, L. Time and length scales in supercooled liquids. *Phys. Rev. E - Stat. Nonlinear, Soft Matter Phys.* **69,** 14–16 (2004).

23. Ruta, B. *et al.* Atomic-Scale Relaxation Dynamics and Aging in a Metallic Glass Probed by X-Ray Photon Correlation Spectroscopy. *Phys. Rev. Lett.* **109,** 1–5 (2012).

24. Berthier, L. Dynamic heterogeneity in amorphous materials. *Physics (College. Park. Md).* **4,** 7 (2011).

25. Ruta, B., Baldi, G., Monaco, G. & Chushkin, Y. Compressed correlation functions and fast aging dynamics in metallic glasses. *J. Chem. Phys.* **138,** (2013).

26. Richter, D., Frick, B. & Farago, B. Neutron-spin-echo investigation on the dynamics of polybutadiene near the glass transition. *Phys. Rev. Lett.* **61,** 2465–2468 (1988).

27. Lunkenheimer, P., Wehn, R., Schneider, U. & Loidl, A. Glassy aging dynamics. *Phys. Rev. Lett.* **95,** 1–4 (2005).

28. Olsen, N. B., Christensen, T. & Dyre, J. C. Time-temperature superposition in viscous liquids. *Phys. Rev. Lett.* **86,** 1271–1274 (2001).

29. Evenson, Z. *et al.* X-Ray Photon Correlation Spectroscopy Reveals Intermittent Aging Dynamics in a Metallic Glass. *Phys. Rev. Lett.* **115,** 175701 (2015).

30. Ruta, B., Giordano, V. M., Erra, L., Liu, C. & Pineda, E. Structural and dynamical properties of Mg65Cu25Y10 metallic glasses studied by in situ high energy X-ray diffraction and time resolved X-ray photon correlation spectroscopy. *J. Alloys Compd.* **615,** 45–50 (2014).

31. Orsi, D. *et al.* Controlling the dynamics of a bidimensional gel above and below its percolation transition. *Phys. Rev. E - Stat. Nonlinear, Soft Matter Phys.* **89,** 1–11 (2014).

32. Bouchaud, J. & Pitard, E. Anomalous dynamical light scattering in soft glassy gels. *Eur. Phys. J. E* **236,** 231–236 (2001).

33. Chaudhuri, P. & Berthier, L. Ultra-long-range dynamic correlations in a microscopic model for aging gels. 1–5 (2016).

34. Ferrero, E. E., Martens, K. & Barrat, J. L. Relaxation in yield stress systems through elastically interacting activated events. *Phys. Rev. Lett.* **113,** 1–5 (2014).

35. Ruta, B., Chushkin, Y. & Monaco, G. Relaxation dynamics and aging in structural glasses. *AIP Conf. Proc.* **1518,** (2013).

36. Angell, C. A., Ngai, K. L., McKenna, G. B., McMillan, P. F. & Martin, S. W. Relaxation in glassforming liquids and amorphous solids. *J. Appl. Phys.* **88,** 3113 (2000).

37. Busch, R., Liu, W. & Johnson, W. L. Thermodynamics and kinetics of the Mg65Cu25Y10 bulk metallic glass forming liquid. *J. Appl. Phys.* **83,** 4134–4141 (1998).

38. Stolpe, M. *et al.* Structural changes during a liquid-liquid transition in the deeply undercooled. *Phys. Rev. B* **93,** 1–7 (2016).

39. Bednarcik, J. *et al.* Thermal expansion of a La-based bulk metallic glass: insight from in situ high-energy x-ray diffraction. *J. Phys. Condens. Matter* **23,** 254204 (2011).

40. Giordano, V. M. & Ruta, B. Unveiling the structural arrangements responsible for the atomic dynamics in metallic glasses during physical aging. *Nat. Commun.* **7,** 10344 (2015).